\documentclass[traditabstract]{aa} 

\usepackage{graphicx}

\usepackage{txfonts}

\usepackage{natbib}

\bibpunct{(}{)}{;}{a}{}{,} 

\begin{document}

   \title{A forming, dust-enshrouded disk at z=0.43: the first example of a massive, late-type spiral rebuilt after a major merger?}


   \author{F. Hammer\inst{1}
 \and
    H. Flores\inst{1}
 \and
 Y. B. Yang\inst{1},
 \and
     E. Athanassoula\inst{2}
 \and
 M. Puech\inst{3,1}
 \and
 M. Rodrigues\inst{1}
     \and 
S. Peirani\inst{4}
         }

   \institute{Laboratoire Galaxies Etoiles Physique et
        Instrumentation, Observatoire de Paris, 5 place Jules Janssen,
        92195 Meudon France\\
              \email{francois.hammer@obspm.fr}
         \and
             Laboratoire d'Astrophysique de Marseille, Observatoire Astronomique de
  Marseille Provence,\\Technopole de l'Etoile - Site de Chateau-Gombert,
  38 rue Fr\'ed\'eric Joliot-Curie, 13388 Marseille C\'edex 13, France
             \and
             ESO, Karl-Schwarzschild-Strasse 2, 85748 Garching bei Munchen, Germany
             \and
             Institut d'Astrophysique de Paris
             }

\date{Received July 1st, 2008; Accepted: November 27th, 2008}



  \abstract{ By combining Ultra Deep Field imagery from HST/ACS with kinematics
  from VLT/GIRAFFE, we derive a physical model of distant galaxies
  in a way similar to that achievable for nearby galaxies.
   A significant part of the evolution in the density of  cosmic
    star formation is related to the rapid evolution of both Luminous IR galaxies (LIRGs) 
  and Luminous Compact galaxies: here we study the properties of a
  distant, compact galaxy, J033245.11-274724.0, which is also a
  LIRG. Given the photometric and spectrophotometric accuracies of data of
  all wavelengths, we can decompose the galaxy into sub-components
  and correct them for reddening.  Combination of deep imagery and
  kinematics provides a reasonable physical model of the galaxy. 
The galaxy is dominated by a dust-enshrouded disk revealed by UDF imagery. 
The disk radius is half
   that of the Milky Way and the galaxy forms stars at a rate of 20
   M$_{\odot}$/yr.  Morphology and kinematics show that both gas and stars
   spiral inwards rapidly to feed the disk and the
   central regions. A combined system of a bar and two nonrotating
   spiral arms regulates the material accretion, induces high
   velocity dispersions, with $\sigma$ larger than 100 km/s and
   redistributes the angular momentum.  The detailed physical
   properties of J033245.11-274724.0 resemble the expectations of
   modeling the merger of two equal-mass, gaseous-rich galaxies, 0.5 Gyr
   after the merger. They cannot be reproduced by any combination of
   intrinsic disk perturbations alone, given the absence of any
   significant outflow mechanisms.  In its later evolution,
   J033245.11-274724.0 could become a massive, late-type spiral that evolves to become part of
   the Tully-Fisher relation, with an angular momentum induced mostly 
   by the orbital angular-momentum of the merger.  


               }

\authorrunning{F. Hammer et al.} 
\titlerunning{a massive, late-type spiral rebuilt after a major merger}

   \maketitle

%


\section{Introduction}

Disk galaxies constitute the majority of the galaxy population
observed in the local universe. They represent 70\% of intermediate-mass
 galaxies (stellar masses ranging from 3 $\times$ 10$^{10}$ to 3
$\times$ 10$^{11}$ M$_{\odot}$), which themselves include at least
two-thirds of the present-day stellar mass density (e.g.,
\citealt{Hammer05}). A model of disk formation developed by 
Fall and Efstathiou (1980), assumed that formation results from
gas cooling and condensation in dark haloes. Their angular momentum
would have been acquired during early interactions \citep{Mo98} and
later the disks may have grown smoothly by gas accretion. Such a
model successfully reproduces flat rotation curves but faces a
difficulty in reproducing the angular momentum of galactic disks:
simulated galaxies have too small angular momentum (e.g., \citealt{Steinmetz99}). However, the
 secular model of disk formation is consistent with observational studies of the Milky Way. 
 It originates from
early studies of the Milky Way that developed into a general description
of the formation of a disk galaxy embedded in a halo (e.g., \citealt{ELS62}). On the other hand, the peculiar nature
of the Galaxy has been questioned \citep{Hammer07}: compared to other
spirals of similar rotational velocity, it exhibits too small stellar
mass, angular momentum, disk radius and halo stars [Fe/H] ratio.
 The Milky Way is possibly peculiar because it has not experienced any
significant collisions since 10-11 Gyrs, in contrast to M31 (\citealt{Ibata05,Brown08}), whose properties are far more typical
(\citealt{Hammer07}). Thus, it could be useful to re-evaluate the
secular model of disk formation.

An important debate is how
to reconcile the presence of numerous present-day thin disks with their fragility
to collisions (\citealt{Toth92}).
All recent studies now appear to
agree that a significant fraction ($\sim$40-90\%)
of local intermediate-mass galaxies
have experienced a major merger within the past 8 Gyrs
(\citealt{Rawat08}, \citealt{Ryan08}, \citealt{Lotz08a}, \citealt{Bell06}). Studies of the merger-rate evolution have now
reached sufficient maturity: the pair fraction of galaxies at
z$\sim$0.6 with a mass ratio ranging from 1:1 to 1:4 is
5$\pm$1\%, regardless of the adopted observational methodology (e.g. \citealt{Bell06}, and references therein). The main uncertainty presently exists in the modelling, i.e. the assumed characteristic time for a true pair
to actually merge, but there are few doubts that major mergers play an important role in shaping  present-day galaxies and thus, the Hubble sequence. 

The above considerations may support the \cite{Hammer05} proposition that disks have been rebuilt
following the encounters of gas-rich spirals. This proposition was
inspired by the remarkable coincidence between the increase with
redshift, up to z=1, in the fraction of actively
star-forming galaxies (including LIRGs, i.e., Luminous InfraRed
Galaxies) and the fraction of galaxies with anomalous
morphologies, including the high fraction of luminous compact  galaxies at z$\geq$0.4, see \cite{Hammer01,Hammer05}. This proposition was followed
by simulations of gas-rich mergers \citep{Robertson06,Governato07},
which demonstrated  that a disk may be re-built after a
merger. Such a conclusion had been already reached by
\cite{Barnes02}, who predicted: "These disks, if subsequently converted
to stars, would be fairly hard to detect photometrically unless viewed
from a favourable orientation." More recently \cite{Lotz08b} 
analysed a large suite of simulated equal-mass, gas-rich mergers
and described most merger remnants as disc-like and dusty. In these simulations, significant disks may be rebuilt only if the encounters are very gas rich, with gas fraction of 50\% or more. The importance of the evolution in the gas content of galaxies is still unknown because direct observations are out of reach before SKA. However, indirect observations based on the O/H abundances of the gas phases of galaxies suggest high gas fractions at z$\sim$ 2 (\citealt{Erb06}), and even at z$\sim$0.6 (\citealt{Liang06,Rodrigues08}).

Could we observe  rebuilt 
disks and what 
 would be
their properties? If most of the mergers occur at z=0.8-2 (e.g.,
\citealt{Ryan08}), a quite significant population
of merger remnants should exist at z=0.6 or even at z=0.4: the typical
remnant timescale is found to be $\sim$2 Gyrs, according to numerical
simulations (e.g., \citealt{Robertson06}), far longer than the characteristic time for a true pair
to merge. As they should form stars rapidly, they are probably dusty
and thus difficult to detect with moderately deep imagery. The optical
disk of the Milky Way (defined to be 3.2 times the disk scale-length)
would be detected out to z=0.4-0.5 (redshifted $R_{25}$ with S/N$\ge$3
in a 1.5 resolution element) using HST/ACS imagery from GOODS. However, this does not apply for a disk enshrouded by dust, which can be
detected only by far deeper exposures from the UDF. The VLT Large
Program IMAGES (\citealt{Yang08}) is observing a
representative sample of $\sim$ 100 z=0.4-0.75 intermediate-mass
galaxies with $M_{J}$(AB)
$\le$ -20.3. Four of these galaxies are located in the Hubble Ultra Deep Field,
and for which we have the ultra-deep imagery from B to H
band and kinematical data (from VLT/GIRAFFE), mid-IR data (from Spitzer)
and deep spectroscopy (from VLT/FORS2). The HUDF galaxy analysed here
is J033245.11-274724.0, at z=0.43462. In Sect. 2, we describe its overall
properties, in Sect. 3 we present the data analysis and modelling, and in Sect.
4, we present our conclusions about the nature of this object. Throughout the
paper, we adopt $H_0=70$ km/s/Mpc, $\Omega _M=0.3$, $\Omega
_\Lambda=0.7$, and the $AB$ magnitude system.


\section{Properties of J033245.11-274724.0 at z=0.43462}

 J033245.11-274724.0 is a very compact galaxy with $r_{half}$=1.7 kpc.
 It was firstly classified as a Tadpole due to its bright core and
 apparent coma-like shape \citep{Neichel08}, on the basis of
 imaging data from GOODS. It is a luminous, massive galaxy with a
 stellar mass comparable or higher than that of the Milky Way. These
 compact objects have been already identified (Guzman et al. 1997;
 Hammer et al. 2001) to be quite numerous at intermediate redshift and
 far more abundant than in the present-day Universe (e.g. Rawat et
 al. 2007). Since they are often strong star-forming objects, their
 emergence at intermediate redshift contributes to the overall decline of the star
 formation density (e.g. Guzman et al. 1997).

  Table 1 describes the overall properties of the galaxy, including photometric data,
  morphological parameters, kinematical data and long-slit spectral measurements.
  Photometric, kinematical and spectroscopic measurements are provided
  by the IMAGES database and can be retrieved from \cite{Yang08},
  \cite{Neichel08}, \cite{Puech08} and \cite{Rodrigues08}. We note that
  the kinematics of compact galaxies such as J033245.11-274724.0 are
  affected by the large pixel size of GIRAFFE, and simulations of
  corresponding data-cubes reveal that large correction factors
  (1.5-2, see \citealt{Puech06} and Figure 5 of \citealt{Puech08}) may
  be needed to recover the flat part of the rotation curve: this
  is due to the fact that the relatively coarse spatial resolution of
  GIRAFFE observations results in a smearing of the velocity gradient
  within each pixel.

J033245.11-274724.0 is an LIRG producing an enormous amount of stars,
given its small size. We can estimate the global extinction of the
system. Using the comparison of $SFR_{IR}$ with SFRs from $H\beta$ and
$H\alpha$ provides $c_{extinction}$= 1.0 for both Orion and standard
extinction curves. This value appears to be robust, since it coincides with
the value obtained from $H\alpha$/$H\beta$ and $H\beta$/$H\gamma$
ratios. Using the ratio of SFRs estimated at $2800\AA$ to SFR in the
infrared, we find $c_{extinction}$=0.54 and 0.79, for the standard and
the Orion extinction curves, respectively. This may favour the Orion
extinction curve, and the contamination of the 2800\AA~ continuum by an
intermediate-age population may explain the differences between the extinction
estimates. The overall galaxy color is $(b-z)_{AB}$=2.28, which at
this redshift corresponds to that of a Sbc galaxy (see Fig. 8 of
\citealt{Neichel08}). Applying the Orion extinction curve to the
overall color of the galaxy resolves most of the discrepancy due to the gap between
Sbc and pure starburst colours ($(b-z)_{AB}$= 0.78). A similar
exercise assuming the standard extinction curve would provide a
dereddened colour bluer than that of a pure starburst, which is not
plausible. In the following, we adopt an Orion extinction curve
with $c_{extinction}$= 1.0.

\begin{table}

\caption{Properties of J033245.11-274724.0}             


{\scriptsize

\begin{tabular}{llllllll}\hline

\multicolumn{8}{l}{Multi wavelength photometry, stellar mass and SFR}\\\hline

 M$_{2800}$ & M$_{B}$ & M$_{V}$ & M$_{K}$&$M_{stellar}$ &  SFR$_{2800}$ & $f_{24\mu}$  &SFR$_{IR}$ \\

 & &  & &log(M$_{\odot}$) & M$_{\odot}$/yr & mJy &M$_{\odot}$/yr   \\

-18.71 & -20.12 & -20.76 &-22.05 &10.8 & 1.9 & 0.353 &21.7 \\\hline

\multicolumn{8}{l}{Morphology: evaluated in observed z band (equivalent to rest-frame R band)}\\\hline

r$_{half}$ & B/T & D/T & $r_{bulge}$ & $r_{disk}$ &$\mu_{0,disk}$ & $incl_{disk}$ & $PA_{disk}$   \\

kpc &   & & kpc & kpc & $mag/"^{2}$ & deg. & deg.    \\

1.7 & 0.14 & 0.51 & 0.13 &1.45 & 20.4 & 39 & -66   \\\hline

\multicolumn{8}{l}{Kinematics: from GIRAFFE measurements \citep{Puech08,Yang08}}\\\hline

V$_{flat}$ & $\sigma$(disk) &v/$\sigma$ &$j_{disk}$ & PA& & &  \\

km/s & km/s &  & km/s$\times$kpc &deg. &  &  &    \\

290 & 74 & 4.0 & 841 & 70 &  &  &    \\\hline

\multicolumn{8}{l}{Spectroscopy}\\\hline

$W_0$(OII) & $H\alpha/H\beta$ &$c_{ext.}$ &$R_{23}$ & O/H &OI &NII & $SFR_{H\alpha}$   \\

\AA &   &  &  & & /$H\alpha$ & /$H\alpha$ &M$_{\odot}$/yr   \\

15 & 6.2 & 1.0 & 1.9 & 9.1 & 0.04 & 0.4 &  3.15   \\\hline

\end{tabular}

}

\end{table} 

  Figure 1 compares the GOODS and UDF images and illustrates how
  vital the depth of UDF is 
for identifying the faintest
  structures in a dust-enshrouded galaxy. The global appearance of the
  galaxy consists of a central bulge with a small bar prolonged by very
  luminous arms, which are surrounded by a red disk. We used GALFIT to
  decompose the z-band galaxy luminosity profile (approximately
  rest-frame R) in individual components. The result is illustrated in
  Fig. 2. Given the appearance of the galaxy, we adopted 
a  3 component fitting (bulge, disk and arms) as provided by
  GALFIT. During our analysis, we attempted to vary
  some parameters such as the bulge Sersic index or its axis ratio.
  All of our attempts confirmed the numbers displayed in Table 1, namely
  the B/T and D/T ratio, disk and bulge radii, and disk inclination and
  PA. The smallest residual is obtained for a low value of the bulge
  Sersic index (n=0.36), but it assumes a very inclined system
  (b/a=0.07) a result which is 
likely affected by the bar. By assuming
  a round bulge we derived a Sersic index of n=3.89, in good agreement with expectations
  for a classical bulge. We also attempted to vary the Sersic
  index of the disk, which provided a best fit solution for n=0.97 in good
  agreement with expectations for a exponential disk. For reason
  of simplicity, we then assumed values of n=4 and 1 for deriving the
  luminosity profile of the bulge and the disk, respectively (Fig. 2).

\begin{figure}

   \centering

\includegraphics[width=8.5cm]{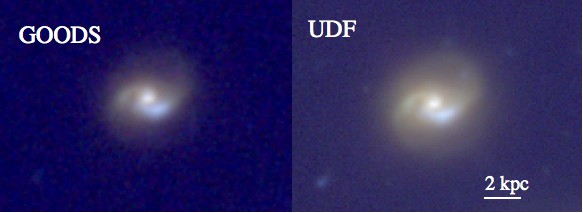}

   \caption{(b, v+i, z ) colour-combined image of J033245.11-274724.0
   from GOODS (left) and UDF (right), assuming the same S/N background
   cut-off (S/N$>$3). Only the depth of the UDF imagery allows us to
   derive robust parameters for the disk (P.A., inclination, size):
   even at z=0.4, dusty disks require extremely deep exposures for
   such measurements. Notice the two small arms surrounding the bar
   near the centre that show a strong color difference suggesting the
    important role of dust extinction. The Figure in the
   electronic version has a higher contrast. }

              \label{Fig1}

    \end{figure}

\begin{figure}
   \centering
\includegraphics[width=8.5cm]{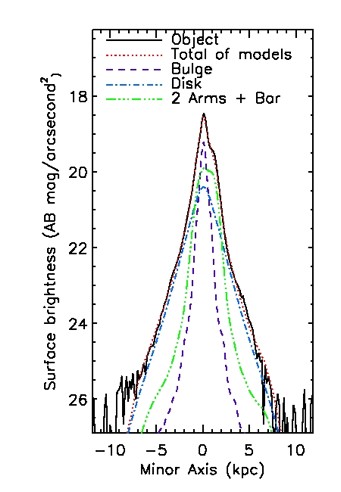}
   \caption{Luminosity profile along the minor axis from the UDF z band image showing the relative contributions of the bulge, disk and barred-arms. This profile was obtained by assuming a round bulge with a Sersic index of 4 and a disk with a Sersic index of 1.}
              \label{Fig1}
    \end{figure}

 The UDF imagery (Figure 1) reveals a red disk surrounding a blue
 bulge, which is quite unusual. The disk colour ranges from (b-z)= 2.8
 to 3.5, i.e. close to that of very old stellar populations. Thus, it is
 likely that the disk of the LIRG is heavily extincted. Indeed, while
 the bar is clearly identifiable between the bluest arm and the bulge,
 it vanishes on the other side. The dust within the disk may also be
 responsible for 
the colour asymmetry of both bar and arms, such that the
 bluest arm is in the foreground to the disk, while the other is in background.
 It suggests another oddity of J033245.11-274724.0: the arms -- as
 well as the bar -- may not be located fully inside the disk, so that a
 significant part of them emerges above and below the disk. Both arms
 have clear extensions that surround the disk, suggesting spiraling
 motions from the disk edge to the centre (see Fig. 1). Gas motions
 are quite complex, since the main disk photometric axis is severely offset from
 the dynamical axis (see Table 1).

We note that J033245.11-274724.0 has been classified by \citep{Yang08} as a rotating basis on the sole basis of its kinematics, in absence of a clear detection of the dust-enshrouded disk (see Fig. 1). The detection of the disk with UDF imaging challenges this classification because of the misalignment between optical and dynamical axes. This also explains why several quantities in Table 1
 differ from those estimated earlier: in this study, the B/T ratio and the disk
inclination (and thus derivation of the velocity) were calculated 
  after applying GALFIT to UDF rather than to GOODS imagery.

\section{Physical model of J033245.11-274724.0}

\subsection{De-reddened properties reveal a dominant dust-enshrouded disk}

  The bulge has a colour intermediate between a Sbc and a late-type
  galaxy, and 
the bluest arm has a colour similar to that of a late
  type galaxy. The oldest stars in the system are likely to inhabit
  the bulge, and it is reasonable to assume that both the bulge and the
  bluest arm are almost not affected by dust. By using NICMOS
  observations in the J and H band (rest-frame near IR), we find that
  the B/T ratio does not vary significantly from our fiducial value
  (0.14; see Table 1), meaning that there is no significant dust-enshrouded component in the bulge. Table 2 presents the
  decomposition of the galaxy into multiple components assuming that the
  extinction affects mostly the red disk component as well as the
  background arm. In this model, the disk dominates
  J033245.11-274724.0, representing 81\% of its stellar mass and more
  than 90\% of its star formation. The appearance of the dereddened
  galaxy would be a disk/arm system with the colour of a late-type
  galaxy (b-z=1.5-1.6) surrounding a bulge with a colour (b-z=2)
  intermediate between a Sbc and a late-type spiral. Assuming a
  uniform colour for all components would not affect our conclusion
  that J033245.11-274724.0 is heavily dominated by its disk, both in terms of
  mass and star formation. We note that the corrected stellar mass
  of J033245.11-274724.0 was estimated from the rest-frame K-band
  magnitude (after correction for extinction) and was assumed the
  extinction-corrected B-V colour in estimating the
  $M_{stellar}$/$L_K$ ratio (see \citealt{Hammer05}).

\begin{table}

\caption{Decomposition into multiple components}             


{\scriptsize

\begin{tabular}{llllllll}\hline

\multicolumn{8}{l}{ observed magnitudes - extinction (assuming late type disk color) - corrected values}\\\hline


 Component & z & b-z & $c_{ext.}$ & $c_{ext.}$ & $b_{c}$ & $SFR_{c}$  &$M_{stellar}$(c) \\

  & &  & /starburst & assum.  & &    \\

  & mag.& mag. & Or. & & mag. &M$_{\odot}$/yr & log(M$_{\odot}$)   \\

bulge       & 22.5 & 2.0 & 1.13 & 0 & 24.5 &0.3 & 8.9 \\\hline


disk          & 21.13 & 2.9 & 2.15 & 1.28 & 19.94 & 20.1 & 10.52 \\\hline


foreg. arm & 22.0 & 1.6 & 0.76 & 0 & 23.6 & 0.65 &  9.1 \\\hline


backg. arm & 22.74 & 2.3 & 1.41 & 0.65 & 23.6 & 0.65 & 9.1  \\\hline


total           & 20.42 & 2.28 & 1.4 & 1.0 & 19.42 & 21.7 &  10.61 \\\hline

\end{tabular}

}

\end{table}

\subsection{Spectral properties: young and intermediate age stars dominate}

We also obtained deep spectroscopy with FORS2
\citep{Rodrigues08}. After correction for extinction (Figure 3), the
spectrum is dominated by the light of young and intermediate age (A/F) stars, as
indicated by the strong emission lines and strong Balmer absorption
lines ($EQW(H\delta)$=4.5\AA). The line ratios are consistent with those
of HII galaxies. There is some evidence that extremely young and hot stars
are present as indicated by the detection of the He5875\AA~ line, and
possibly Wolf-Rayet features, such as the blue bump and possible
detection of [SiIII]4565]\AA, [NV]4620\AA, [HeII]4686\AA, and
[NII]5720\AA. The global metal abundance is over-solar,
at 12+log(O/H)=9.1, which can be explained by the very large density of
recent and past star formation all over the
galaxy. The presence of
metal absorption lines (CaIIK, G, MgI \& NaD) indicates the presence
of relatively old stars. We modeled the stellar populations using the
STARLIGHT code (e.g. Cid-Fernandes et al., 2008). The high quality of
the FORS2 spectrum is at least comparable to that of a typical galaxy
from the SDSS. It provides a distribution of stellar age including a
large mass-component of young to intermediate age stars (mass scaled:
20\% with age $\le$ 0.3 Gyrs, 33\% with age = 0.5-0.8 Gyrs), the remainder
being associated with older ages ($>$ 2 Gyrs). This confirms the recent
formation of a significant part of the stellar mass in
J033245.11-274724.0.


 \begin{figure}
   \centering
\includegraphics[width=9cm]{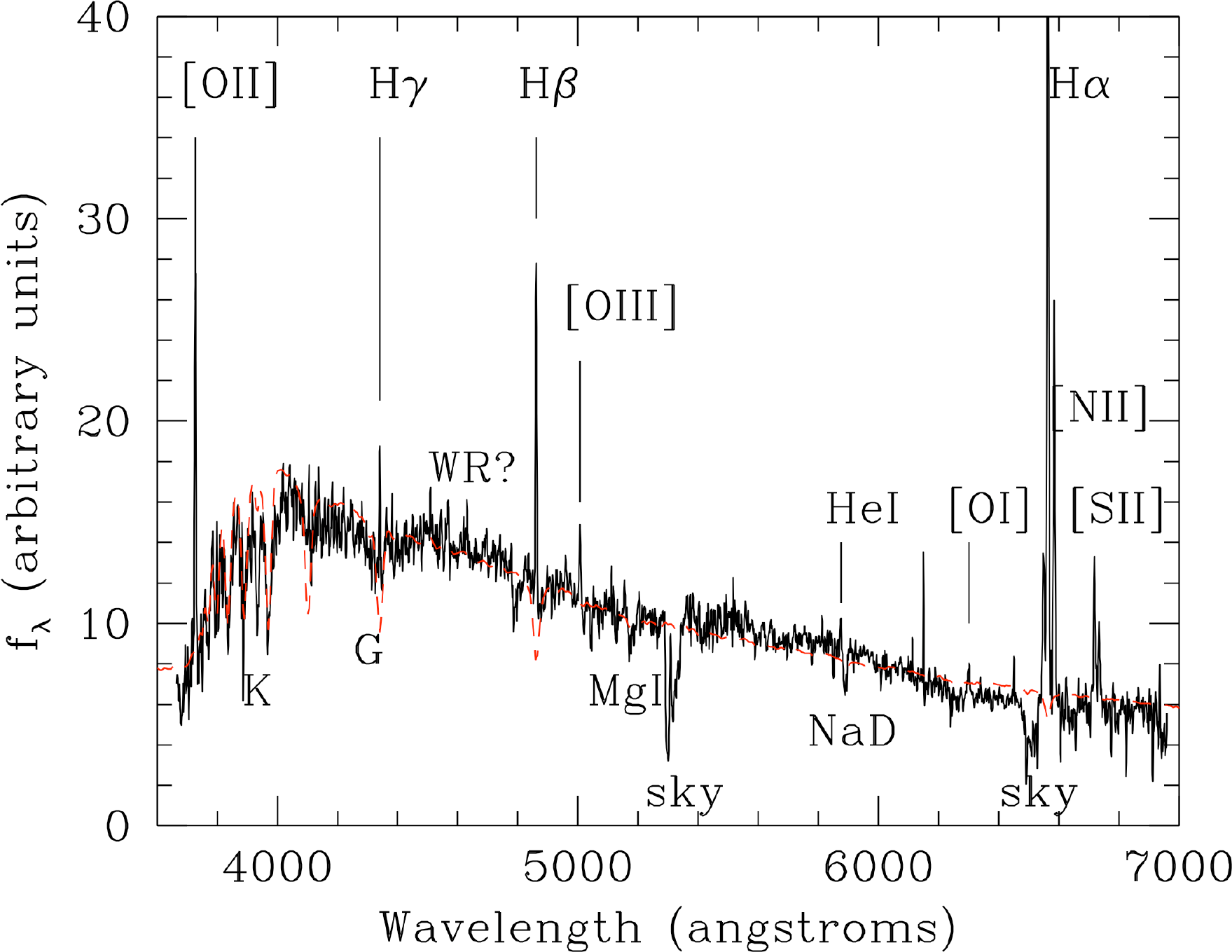}
   \caption{FORS2 spectrum of J033245.11-274724.0 after dereddening the continuum by $c_{extinction}$=1.0 (Orion extinction curve). The spectral energy distribution can be reproduced (red line) by a set of models with stellar ages t from 0.25 to 1.2 Gyr assuming $\tau$ (SFR$\sim$ exp(-t/$\tau$)) from 0 to 5 Gyr. 
}

              \label{Fig3}
    \end{figure}

 \subsection{Morpho-kinematic properties: a hot disk out of equilibrium}

J033245.11-274724.0 is dominated by a dust-enshrouded disk with a
quite peculiar {\it two-armed} system. The kinematical axis is
strongly offset from the photometric major axis of the disk and is
slightly offset from the galaxy centre (Fig. 4, upper panels),
arguing for a non-axisymmetric, heavily perturbed and out of
equilibrium disk. The outskirts of J033245.11-274724.0 exhibit extremely high
velocity dispersions, ranging from 60 to 105 km/s (Fig. 4 and
\citealt{Yang08}). \cite{Puech07} estimated the intrinsic
velocity dispersion of the disk by excluding the central pixels, which
are likely to be affected by the central
velocity gradient linked with the rotation at low spatial resolution. By excluding the
central regions, \cite{Puech07} found $\sigma_{disk}$=74 km/s for
J033245.11-274724.0, a value far larger than expected for a thin
disk. The disk clearly dominates the outskirts of the galaxy
as it can be inferred from the luminosity profile (Fig. 2), and thus, the bulge cannot bring any significant contribution to the high dispersion.

We also note that the velocity dispersion peaks are located near the ends of the arms, which is rather unexpected for the velocity
field of a rotating disk observed at low spatial resolution
\citep{Flores06}. Fig. 4 (bottom-left) shows a sketchy
representation of the galaxy, which reproduces both its morphology and
kinematics.

The disk of J033245.11-274724.0 is rotating too fast for its stellar
mass. Indeed, assuming a disk inclination of 39$^\circ$ (see Table 1), the fitting
of the velocity field \citep{Puech08} provides $V_{flat}$=290km/s,
while at its stellar mass, the Tully-Fisher relation predicts
$V_{flat}$=190km/s at z=0. This shift relative to the Tully-Fisher relation is
significant, and cannot be due to an underestimate of the inclination:
adopting $V_{flat}$=190km/s would infer a disk inclination of
74$^\circ$, which is clearly inconsistent with observations. It is
thus likely that the disk -- which actively forms stars at a rapid
rate -- would increase its stellar mass further, and would evolve into agreement with the local
Tully-Fisher at z=0.

\subsection{Comparison to simulations: a merger remnant?}

 Similar features (bar + {\it two armed} system) are observed in
 current simulations of galaxy mergers well after the collision (see
 Fig. 4, bottom-right). These features do not follow 
the circular
 gas motions\footnote{we encourage the reader to examine the simulation
 at
 http://www.ifa.hawaii.edu/$\sim$barnes/research/gassy\_mergers/index.html,
 model with 1:1 mass ratio, INClined orbit and pericenter $r_p$=0.2.
 }, and their misalignment with respect to the disk may be due to the
 fact that they originate preferentially in one of the two progenitors
 (see \citealt{Barnes02}). Their main role is to funnel gas
 towards the galaxy to fuel the formation of the bulge and the disk. After disk
 stabilisation, they may disappear in the simulation. The global
 resemblance between the simulations and the observations (morphology
 and kinematics) is rather 
striking. N-body/SPH simulations by \cite{Barnes02} use a
 minimum of prescriptions that may simplify this comparison by
 minimising the number of physical ingredients. The predicted
 rebuilt disks in \cite{Barnes02} are indeed warped, and the kinematics of
 J033245.11-274724.0 (kinematical axis strongly offset from the
 photometric axis \& velocity dispersion peaks) is clearly reproduced by simulations. We note that the
 example shown in Fig. 4 illustrates the gas component only. It can,
 nevertheless, be compared to the observed stellar distribution, since
 most of the emission in the J033245.11-274724.0 system should be
 dominated by young and massive stars, of ages relatively small in
 comparison with dynamical timescales.


\begin{figure}
   \centering
\includegraphics[width=9cm]{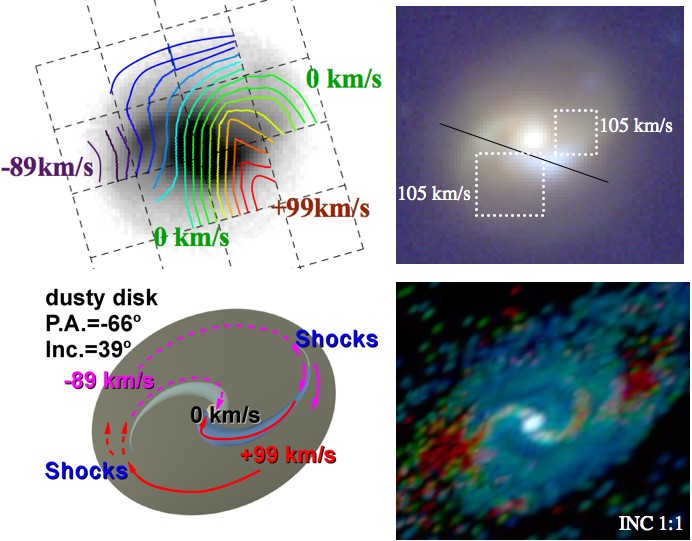}
   \caption{ {\it Upper left:} v-band image of J033245.11-274724.0 on
     which the GIRAFFE iso-velocities have been superimposed. Colour
     coding {\bf is} the same {\bf as} in \cite{Yang08}. {\it Upper
     right:} b-v+i-z colour-combined image of J033245.11-274724.0 on
     which the dynamical axis (full lack line) and the velocity
     dispersion peaks (dotted squared boxes) at 105 km/s have been
     superimposed. The entire galaxy shows large velocity
     dispersion \citep{Yang08}. The Figure in the electronic version
     is of higher contrast. {\it Bottom left:} A sketchy model of the
     galaxy showing the dusty disk, and the gas motions (purple:
     approaching; red: recessing) from the background (dotted lines)
     to the foreground of the disk (full lines). The bar close to the
     centre is inclined with respect to the disk plane, explaining its
     disappearance at the top-right of the bulge as well as the
     asymmetry between the two arms. The resulting dynamical
     axis is thus caused by the gas motion both through the bar and
     around the disk. {\it Bottom right:} result of a galaxy collision
     ( 1:1 mass ratio) with an inclined orbit (close encounters) from
     \cite{Barnes02}, 0.5 Gyr after the merger. The gas wraps up
     inside the disk, creating a bar and a symmetric-arm system whose
     motion does not follow the gas rotation, creating shocks and high
     velocity dispersions (seen as red particles in the simulation),
     which are consistent with observations. }

              \label{Fig2}
    \end{figure}

  \section{Discussion}

The wealth of data for J033245.11-274724.0 has allowed us to identify a
faint, red disk that dominates both its star formation and stellar
mass. The disk is indeed small with a radius half that of the Milky
Way, while having a similar stellar mass and forming stars 20 times
more rapidly. The presence of a bar close to the centre, a symmetrical
two-arm system extending on both sides of the disk plane towards
the disk edge, is indicative of a significant spiraling motion of both
gas and stars from the outskirts to the centre. Our sketchy
representation of J033245.11-274724.0 reproduces well the
morphological features as well as the kinematics, assuming that the
symmetric arms do not follow the circular gas motions, and 
thus create shocks that are revealed by the observed velocity 
dispersion peaks in the 2D velocity dispersion map. 
The overall properties
of J033245.11-274724.0 are thus consistent with the modeling of the
remnant of a merger of two gaseous, equal-mass disks on an inclined
orbit. It appears 
that 
 by combining deep UDF imaging 
with GIRAFFE
integral field spectroscopy, we may have revealed the presence of a dust-enshrouded disk
that is also a merger remnant.

\subsection{A perturbed disk or a merger remnant? }

 Does the success of our physical model in reproducing the observed
 features necessarily imply that J033245.11-274724.0 is a merger
 remnant ? Before reaching this conclusion, we examine some
 possible alternatives. Firstly, the evidence that the system is
 disturbed and not in equilibrium might be explained by a 
passing, tidal
 interaction. However, investigations of the field surrounding
 J033245.11-274724.0 do not reveal any possible companions within
 several hundreds of kpc, using numerous existing spectroscopic or
 photometric redshifts in the field available for all
 galaxies with apparent luminosities down to one third of that of
 J033245.11-274724.0. Examining the UDF imagery in z band, the
 brightest nearby galaxy (with no redshift available) within 50 kpc
 has a z-band luminosity that is 1/15th that of J033245.11-274724.0. Within
 30 kpc, the brightest nearby companion has a z-band luminosity
 that is 1/200th that of J033245.11-274724.0. It is thus very unlikely that
 a fly-by interaction may be the
 cause of the exceptional properties of this galaxy.

Secondly, internal processes related to the starburst may be considered to
explain its properties. In particular, the fact that the
kinematical axis is severely offset from the photometric axis may
be indicative of gas super-winds (outflows) induced by supernova. These
phenomena are observed in many starbursting galaxies (e.g.
\citealt{Heckman00}). We tested such an alternative by comparing
the velocity of the emission-line system with that of the absorption-line system, and by examining the line profile of the NaD absorption
lines, i.e. the two methods outlined by \cite{Heckman00}. By measuring
ten robust emission lines ($H\gamma$, $H\beta$, [OIII]4959 and 5007,
HeI, [OI], [NII]6548 and 6583 and $H\alpha$), we find a remarkable
agreement between those lines with a scatter of
only 15 km/s. We excluded the two [SII] lines, since they exhibit a much
larger shift, presumably due to the fact that they lie at the edge of
the spectrum, where wavelength calibration may have failed. Examining
the absorption lines of a reasonable S/N (Balmer H12 and H11,
CaII K, CaI, NaI-D1 and D2), we find a larger 
scatter of 52 km/s as expected. The absorption and emission
line system appear to have the same velocity, to within an uncertainty of 33 km/s,
a value far below the expected uncertainties dominated by absorption-line measurements (recall that the FWHM spectral resolution is 450
km/s). The contribution of the ISM to the NaD lines may be evaluated
following Heckman \& Lehnert (2000) from the expected relationship
for typical stellar populations: EQW(NaD)=0.75$\times$ EQW(MgIb). For
J033245.11-274724.0, we find EQW(NaD)=2.8\AA~ and EQW(MgIb)=4.0\AA~,
which implies a negligible contribution from the ISM. From both tests,
J033245.11-274724.0 appears to be part of the \cite{Heckman00}
sample of ``strong stellar contamination" starbursts, i.e. those
showing no evidence of outflows. Examination of the emission-line
profiles also show no evidence of an outflow. An absence of a
significant outflow contribution is unsurprising given the high
mass of the galaxy, whose gravitational forces prevent stellar
super-winds dominating the velocity field.

Given the above, we have no suitable explanations of J033245.11-274724.0  other than the merger-remnant hypothesis. Table 3 illustrates this by presenting the two alternatives, a perturbed disk and a merger remnant, and testing them against the main observed properties. 

\begin{table*}

\caption{A comparative test of the perturbed disk and the merger remnant hypotheses}             


{\scriptsize

\begin{tabular}{lll}\hline



 Observed property & best hypothesis & comments \\

\hline\hline

\\

much redder disk than the bulge         & merger remnant & rapid disk rebuilt is predicted (\citealt{Governato07}): renewed disks are thus star forming and dusty (\citealt{Lotz08b})\\\hline
\\

bar \& two arms  system    & both & expected for both (see text) \\\hline

\\

misaligned dynamical axis  & merger remnant & in absence of significant outflow, it reveals a system not at equilibrium, difficult to explain except by a merger
\\\hline

\\

average disk velocity dispersion (74 km/s) & merger remnant & unclear which perturbation other than merging can provide such a hot disk ($\sigma$$>$56km/s) \\\hline

\\

velocity dispersion peak locations & merger remnant & only two minor mergers near the two locations may explain such features in a perturbed disk \\\hline

\\

very high disk star formation density & merger remnant & regulation of star formation between gravitational instabilities and porosity  is expected in rotating disks (e.g. \citealt{Silk97})\\\hline

\\

discrepancy from the Tully Fisher & merger remnant & expected in a merger (e.g. \citealt{Puech07}) and quite unexpected in a perturbed disk (see above)\\\hline

\\

half of the stars with ages lower than 800Myrs & merger remnant & correspond to the merger time-scale  during which these stars are formed\\\hline

\end{tabular}

}

\end{table*} 

   \subsection{A merger remnant of a 1:1 or a 1:3 fusion? }

  Does the success of our physical model necessarily imply that J033245.11-274724.0 is a merger remnant of a 1:1 mass ratio merger? 


 Using simulations by \cite{Barnes02}, we also tested a model
 with 3:1 mass ratio. While this also predicts a shape similar of the
 ``bar plus 2 arms system", it is unable to reproduce both the
 morphology and the high velocity dispersion peaks of
 105 km/s. On the other hand, we tried to reproduce the
 \cite{Barnes02} simulations of 1:1 and 1:3 mass ratio using the same
 parameters but with the Gadget2 code \citep{Springel05}. A bar is
 still present (and persistent), but with a less pronounced S shape.
 Different recipes of gas viscosity and star formation may explain
 these discrepancies. Admittedly, it is beyond the scope of this paper
 to provide an accurate prediction of the mass ratio of the
 progenitors of J033245.11-274724.0, although existing simulations
 clearly favour a range from 1:1 to 1:3.

 It is also important to note that the simulations by
 \cite{Barnes02} predict that a significant fraction of the gas lies
 in the remnant nuclei and not in the disk. On the other hand, in the
 \cite{Lotz08b} realisations several merger remnants have the
 appearance of Sbc galaxies. A full quantitative model of this
 source would be highly desirable to investigate the formation stage
 of the J033245.11-274724.0 disk. For example,
 \cite{Hopkins08} generated a full suite of galaxy-galaxy
 simulations to examine the properties of their corresponding remnants. Their
 simulations may include J033245.11-274724.0 as suggested by their
 abstract: "The efficiency of disk destruction in mergers is a strong
 function of gas content -- our model allows us to predict explicitly
 and demonstrate how, in sufficiently gas-rich mergers (with
 quite general orbital parameters), even 1:1 mass-ratio mergers can
 yield disk-dominated remnants, and more realistic 1:3-1:4 mass-ratio
 major mergers can yield systems with $<$ 20\% of their mass in
 bulges." Motivated by this, we attempted to estimate the gaseous
 content of the galaxy by assuming that it follows the
 Schmidt-Kennicutt relation between star formation and gas
 surface densities (e.g. Kennicutt, 1998). For this, we further
 assumed that the star-formation density could be derived from a surface
  radius of 2 $\times$ $r_{half}$= 3.4 kpc, and that the gas
 surface density could be derived from the maximal extent of the [OII]
 emission detected by GIRAFFE ($r_{OII}$= 5.4 kpc after deconvolution
 from the natural seeing, FWHM=0.8 arcsecond). We then measured a star
 formation surface density of 0.6 M$_{\odot}$$yr^{-1}$$kpc^{-2}$, a
 gas mass density of 2.5 $\times$ $10^{8}$ M$_{\odot}$$kpc^{-2}$ and a
 high total gas mass of 2.3 $10^{10}$M$_{\odot}$. Assuming the
 stellar mass derived in Table 2 (after accounting for dust extinction
 effects), we reached the conclusions that the J033245.11-274724.0
 system may include up to 37\% of gas. Since a significant fraction of
 the stars (approximately half, see Sect. 3.2) may have been formed
 during the interaction, the gas fraction in the progenitors of
 J033245.11-274724.0 may had been much higher, up to 67\%. Such a
 value would be consistent with significant disk rebuilding as demonstrated by
 the \cite{Robertson06} simulations.

We thus are left with an Occam's razor-type argument: the overall and
detailed properties of J033245.11-274724.0 are well reproduced by a
major merger remnant (Table 3), we do expect the existence of such
a remnant from interactions between gas-rich galaxies, and such an event
may be quite common given existing merger statistics, while other explanations
require the combination of unknown mechanisms. We thus conclude that
J033245.11-274724.0 may be the first conclusive example of a rebuilt
disk in the intermediate redshift Universe.

\begin{figure}
   \centering
\includegraphics[width=9cm]{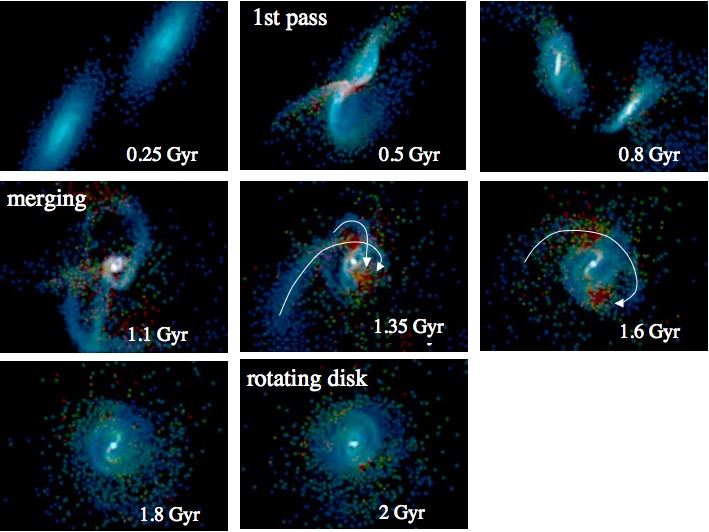}
   \caption{Inclined-orbit model from  \cite{Barnes02} with time scaled for a fusion of two galaxies with half the mass of the Milky Way. The arrows in the central panels (at 1.35 and 1.6 Gyr) indicate the gas motions and the exchanges of angular momentum after the merger. The orbital angular momentum is perpendicular to the observed plane, in contrast to the angular momentum of the progenitors (see panels at 0.25 and 0.8 Gyr), resulting in a face-on rotating disk at the end of the simulation. The bar and  
{\it two armed} system regulate the final disk rotation. Note that the {\it two armed} system does not rotate (compare panels at 1.6 and 1.8 Gyr). }

              \label{Fig4}
    \end{figure}

\section{Concluding remarks}


J033245.11-274724.0 is a warped disk that is dust enshrouded and
forms stars at a very high rate, given its small size. Examination of
the exchanges of angular momentum in \cite{Barnes02} simulation is 
instructive (Fig. 4). Before the second impact, the angular momentum
of the two equal mass progenitors are almost perpendicular to that of
the newly formed disk, whose plane is determined by the orbital
angular momentum. After the second impact, the gas and stars roll up
surrounding the newly formed bulge following trajectories that result
from their initial angular momentum combined with the tidal forces.
The {\it two-armed} system plays the role of the bar,
redistributing the angular momentum of the infalling material towards
the bar and then to the bulge (by means of dispersion) or towards the
newly formed disk. At the stage represented by J033245.11-274724.0, most of the
stars and most of the newly formed stars are located in the forming
disk. This transition is quite long, since the bar plus non rotating {\it
two-armed} system are seen for about 20\% of the total time in the
simulation \citep{Barnes02}. The total time of the simulation is
approximately 2 Gyrs for encounters with half the Milky Way mass,
 the first impact occurring at 0.5 Gyr, the merging of the nuclei
occurring at 1.1 Gyr, the appearance of the non rotating {\it two-armed} system
at 1.6 Gyr and its disappearance at 1.9 Gyr. These numbers are quite
consistent with the ages of the stellar populations that dominate the
J033245.11-274724.0 continuum. On the other hand, the creation of a disk with
3.7 $\times$ $10^{10}$M$_{\odot}$ would require 1.8 Gyr at the rate of
20 M$_{\odot}$/yr (see Table 2), i.e. a time that is long compared to
the time it takes the disk to settle after the merger in the
simulation ($<$0.5 Gyr). The observed stellar mass in the
J033245.11-274724.0 disk of course consists of young stars, although a
significant part consists of intermediate-age stars, probably produced
during the first stages of the merger, and possibly of older stars
accreted from the progenitors. The colour of the dereddened disk
(late type galaxy) is consistent with such a mix of stellar
population.

What  will the final product of J033245.11-274724.0 be? Since most of the star formation has been induced in the disk, one could assume a progressive inside-out disk-building from the residual material accreting onto the disk. As the infalling gas becomes rarer, the star formation and the dust production will decrease and the galaxy will take the appearance of a blue star forming disk. In its further evolution to reach the local Tully-Fisher relation, J033245.11-274724.0 needs to form a larger and more massive disk, which may be followed by  bulge growth by gas accretion from the residual bar. Assuming that the 37\% residual gas fraction estimated in Sect. 4.2 will have been transformed into stars by z=0,   the galaxy may reach the local Tully Fisher relationship. The observed bar in J033245.11-274724.0 is very small (and blue, see Fig. 1) and the simulation does predict a residual small bar a few tenths of Gyrs later. Our observations show that most of the stellar mass is in the disk: it is probable that the newly formed system would have a bulge to total ratio ranging from 0.03 to 0.09 (e.g. Sc or later type), depending on whether the material in the {\it two-armed} system will be deposited in the disk or in the bulge, respectively. Thus, our observations and modeling of J033245.11-274724.0 are highly indicative of the formation of a massive, late-type spiral whose angular momentum support is mostly provided by the orbital momentum of the last merger.

 It is quite a challenge to find similar examples in the local Universe, which is moreover lacking LIRGs (0.5\%), compact luminous galaxies (a few percents, \cite{Rawat07}) and, overall, blue bulges  at the centre of extremely red disks. NGC 1614 is a powerful LIRG and a merger remnant and shows some similarities with J033245.11-274724.0. However, in NGC 1614  most of the starburst activity is in the core \citep{Alonso01}, which is probably induced by the recent coalescence of the two progenitor nuclei (\citealt{Neff90}, and see Fig. 4, panel at 1.1Gyr): it seems to be in a less advanced merger stage than J033245.11-274724.0.

Finally, the occurrence of these merger remnants in the past may be far more common than today. Observed at relatively shallow depth (compared to UDF), J033245.11-274724.0 appears to be a Tadpole due to its bright blue arm. Tadpoles represents 2-5\% of the massive galaxies at z=0.6-1 \citep{Brinchmann98,Neichel08}, and some might be similar to J033245.11-274724.0, possibly seen at different viewing angles. In the near future, we may be able to evaluate whether J033245.11-274724.0 is an isolated case or a more general phenomenon of disk rebuilding. 


\begin{acknowledgements}

We are especially indebted to Josh Barnes for his comments and criticisms in reading an early version of the manuscript as well as for making publicly his simulations of galaxy mergers on his web pages. We are very grateful to the referee whose suggestions have helped a lot to improve the manuscript.

 \end{acknowledgements}

\end{document}